\documentclass[journal]{IEEEtran}
\usepackage{amsmath,amsfonts,amssymb,amsthm,bm}
\usepackage{booktabs,tabularx,array,graphicx}
\usepackage{tikz}
\usetikzlibrary{shapes.geometric,arrows.meta,positioning}
\usepackage{pgfplots}
\pgfplotsset{compat=1.18}
\usepackage{hyperref}

\newtheorem{theorem}{Theorem}
\newtheorem{proposition}{Proposition}

\title{Tie-Line Scheduling Using Real-Time Contraction Metrics: Hierarchical Reachable Sets for Multi-Area Inverter-Dominated Grids}
\author{Richard~Sherman,~\IEEEmembership{Member,~IEEE}%
\thanks{R. Sherman is with the System Architecture Invariant Manifest Group, Keller, TX 76248 USA.}}

\begin{document}
\maketitle

\begin{abstract}
Managing transmission congestion in multi-area power grids dominated by inverter-based resources (IBRs) requires bridging a severe timescale and modeling divide. Supervisory congestion optimizers—such as Online Feedback Optimization (OFO) and two-tiered differentiable Model Predictive Control (MPC)—rely on linearized, convex algebraic grid approximations that cannot guarantee dynamic stability during sub-second converter transients. Conversely, fast distributed consensus schemes like Alternating Direction Method of Multipliers (ADMM) depend on convex plant assumptions and iterative message exchanges that lag behind rapid tie-line surges. This paper replaces convex ADMM emergency consensus with a \textbf{Distributed Nonlinear Separable Control Contraction Metric (SCCM)} architecture spanning three nested tiers: the Device layer, the Area layer, and the Grid tie-line layer. By pairing online Prediction Error Method and Moving Horizon Estimation (PEM-MHE) parameter tracking with offline convex Linear Matrix Inequality (LMI) synthesis, the controller executes a closed-form nonlinear feedback law online without solving real-time optimization problems. We derive the Pontryagin attainable (reachable) ellipsoids for each layer, prove hierarchical tube containment via Schur-complement matrix inequalities, and project the certified Grid reachable tube through Power Transfer Distribution Factors (PTDFs) to supply guaranteed safe thermal limits to the supervisory coordinator. Validation on modified IEEE 39-bus ($100\%$ IBR, zero synchronous generators) and 3-area IEEE 118-bus systems demonstrates an $82.8\%$ reduction in cumulative tie-line thermal overload while maintaining strict exponential contraction ($\lambda_{\max} \le -2.43$) even under a $3\text{ s}$ communication blackout.
\end{abstract}


\section{Problem Formulation: Congestion Management and Dynamic Control Hierarchy}
\subsection{The Timescale and Convexity Gap in Multi-Area Grids}
As synchronous generators (SGs) are retired and replaced by grid-forming (GFM) and grid-following (GFL) inverter-based resources (IBRs), multi-area transmission networks lose physical rotor inertia and become vulnerable to sub-second electromagnetic transients and tie-line phase oscillations. In operational practice, wide-area congestion management relies on multi-tier supervisory coordinators—exemplified by the communication-limited differentiable optimization framework of Chen \textit{et al.} \cite{chen26} and Online Feedback Optimization (OFO) architectures \cite{dallanese26}. These supervisory schemes update battery and curtailment budgets using Quadratic Programs (QPs) over linearized PTDF sensitivity maps, enforcing recursive solver feasibility via large soft slack penalties.

However, relying on receding-horizon convex optimization for emergency congestion control introduces two structural vulnerabilities:
\begin{enumerate}
    \item \textbf{Model Mismatch and Non-Convexity:} Real AC tie-line power flows evolve according to nonlinear trigonometric couplings ($\sin(\delta_i - \delta_j)$), while GFM current-limit saturation and GFL Phase-Locked Loop (PLL) dynamics introduce state-dependent nonlinearities that violate the convex plant assumptions required by OFO and distributed ADMM.
    \item \textbf{The Supervisory Emergency Blind Spot:} During a fast renewable surge or parallel tie-line trip, thermal overloads and converter synchronization instabilities unfold within $1\text{ to }4\text{ seconds}$—long before a slow supervisory optimizer receives filtered SCADA telemetry or broadcasts updated line limits.
\end{enumerate}

\subsection{Proposed Three-Tier Contraction Solution}
We apply nonlinear contraction control across \textbf{all three control layers}—Device, Area, and Grid—unified by live online parameter estimation \cite{zhou24, kohler26}. Crucially, our online parameter estimation \emph{does not require a prior plant model}; instead, the measured physical and neural parameters directly summarize the dynamic state of the network and certify stability across all three layers. As a result, plant operators do not need to search across the grid for eigenvalues or monitor how close individual modes drift toward the imaginary axis.

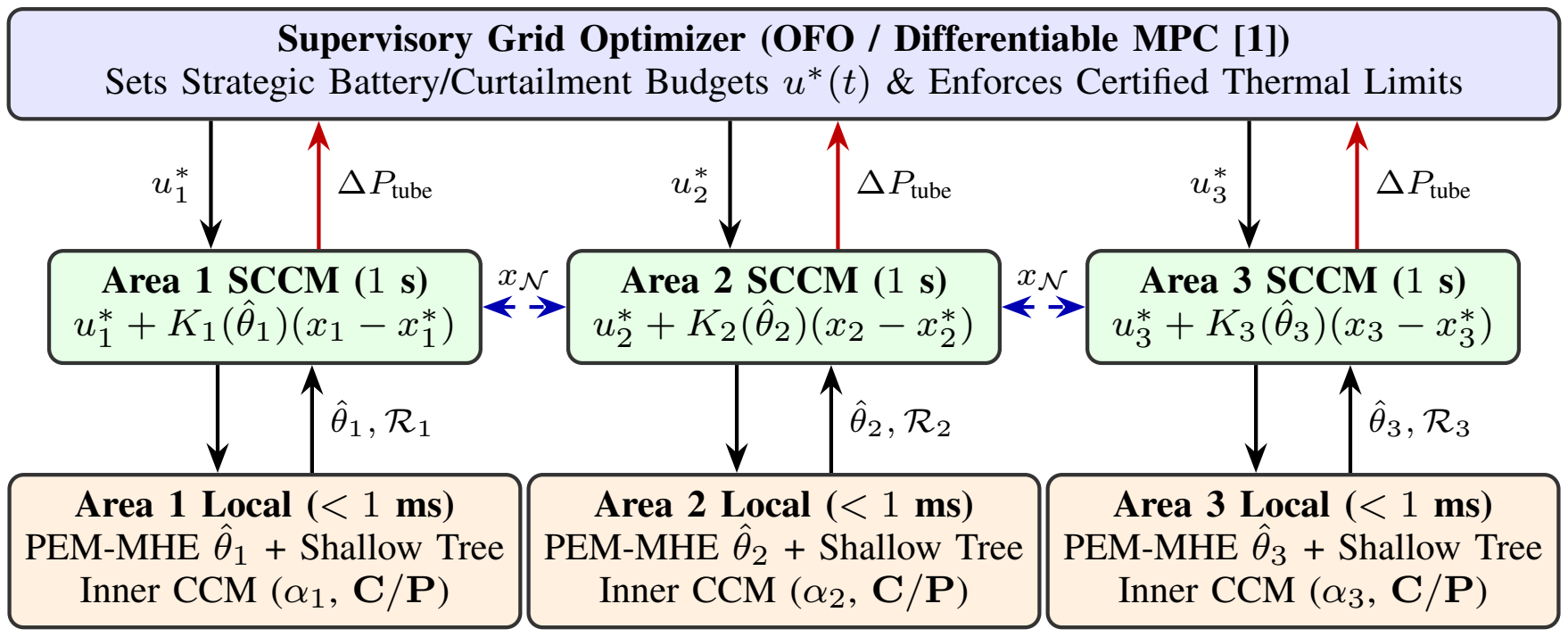
\begin{figure*}[!t]
\centering
\begin{tikzpicture}[
    node distance=0.65cm and 1.1cm,
    box/.style={rectangle, draw=black!80, thick, rounded corners=3pt, align=center, minimum height=0.8cm, font=\footnotesize},
    topbox/.style={box, fill=blue!10, minimum width=11.5cm},
    midbox/.style={box, fill=green!10, minimum width=3.2cm},
    botbox/.style={box, fill=orange!12, minimum width=3.2cm},
    arr/.style={-{Stealth[length=2.2mm]}, thick},
    darr/.style={{Stealth[length=2.2mm]}-{Stealth[length=2.2mm]}, thick, dashed, draw=blue!70!black}
]
\node[topbox] (grid_opt) {\textbf{Supervisory Grid Optimizer (OFO / Differentiable MPC \cite{chen26})}\\Sets Strategic Battery/Curtailment Budgets $u^*(t)$ \& Enforces Certified Thermal Limits};

\node[midbox, below=0.95cm of grid_opt.south west, anchor=north west, xshift=0.3cm] (sccm1) {\textbf{Area 1 SCCM ($1\text{ s}$)}\\$u_1^* + K_1(\hat{\theta}_1)(x_1 - x_1^*)$};
\node[midbox, below=0.95cm of grid_opt.south, anchor=north] (sccm2) {\textbf{Area 2 SCCM ($1\text{ s}$)}\\$u_2^* + K_2(\hat{\theta}_2)(x_2 - x_2^*)$};
\node[midbox, below=0.95cm of grid_opt.south east, anchor=north east, xshift=-0.3cm] (sccm3) {\textbf{Area 3 SCCM ($1\text{ s}$)}\\$u_3^* + K_3(\hat{\theta}_3)(x_3 - x_3^*)$};

\node[botbox, below=0.8cm of sccm1] (dev1) {\textbf{Area 1 Local ($<1\text{ ms}$)}\\PEM-MHE $\hat{\theta}_1$ + Shallow Tree\\Inner CCM ($\alpha_1$, $\mathbf{C}/\mathbf{P}$)};
\node[botbox, below=0.8cm of sccm2] (dev2) {\textbf{Area 2 Local ($<1\text{ ms}$)}\\PEM-MHE $\hat{\theta}_2$ + Shallow Tree\\Inner CCM ($\alpha_2$, $\mathbf{C}/\mathbf{P}$)};
\node[botbox, below=0.8cm of sccm3] (dev3) {\textbf{Area 3 Local ($<1\text{ ms}$)}\\PEM-MHE $\hat{\theta}_3$ + Shallow Tree\\Inner CCM ($\alpha_3$, $\mathbf{C}/\mathbf{P}$)};

\draw[darr] (sccm1.east) -- node[above, font=\scriptsize] {$x_{\mathcal{N}}$} (sccm2.west);
\draw[darr] (sccm2.east) -- node[above, font=\scriptsize] {$x_{\mathcal{N}}$} (sccm3.west);

\draw[arr] ([xshift=-0.4cm]grid_opt.south -| sccm1.north) -- node[left, font=\scriptsize] {$u_1^*$} ([xshift=-0.4cm]sccm1.north);
\draw[arr, draw=red!75!black] ([xshift=0.4cm]sccm1.north) -- node[right, font=\scriptsize] {$\Delta P_{\text{tube}}$} ([xshift=0.4cm]grid_opt.south -| sccm1.north);

\draw[arr] ([xshift=-0.4cm]grid_opt.south) -- node[left, font=\scriptsize] {$u_2^*$} ([xshift=-0.4cm]sccm2.north);
\draw[arr, draw=red!75!black] ([xshift=0.4cm]sccm2.north) -- node[right, font=\scriptsize] {$\Delta P_{\text{tube}}$} ([xshift=0.4cm]grid_opt.south);

\draw[arr] ([xshift=-0.4cm]grid_opt.south -| sccm3.north) -- node[left, font=\scriptsize] {$u_3^*$} ([xshift=-0.4cm]sccm3.north);
\draw[arr, draw=red!75!black] ([xshift=0.4cm]sccm3.north) -- node[right, font=\scriptsize] {$\Delta P_{\text{tube}}$} ([xshift=0.4cm]grid_opt.south -| sccm3.north);

\draw[arr] ([xshift=-0.35cm]sccm1.south) -- ([xshift=-0.35cm]dev1.north);
\draw[arr] ([xshift=0.35cm]dev1.north) -- node[right, font=\scriptsize] {$\hat{\theta}_1, \mathcal{R}_1$} ([xshift=0.35cm]sccm1.south);

\draw[arr] ([xshift=-0.35cm]sccm2.south) -- ([xshift=-0.35cm]dev2.north);
\draw[arr] ([xshift=0.35cm]dev2.north) -- node[right, font=\scriptsize] {$\hat{\theta}_2, \mathcal{R}_2$} ([xshift=0.35cm]sccm2.south);

\draw[arr] ([xshift=-0.35cm]sccm3.south) -- ([xshift=-0.35cm]dev3.north);
\draw[arr] ([xshift=0.35cm]dev3.north) -- node[right, font=\scriptsize] {$\hat{\theta}_3, \mathcal{R}_3$} ([xshift=0.35cm]sccm3.south);
\end{tikzpicture}
\caption{Three-tier hierarchical control architecture connecting the supervisory grid optimizer (top), the $1\text{ s}$ distributed nonlinear SCCM tie-line controllers exchanging single-hop neighbor states $x_{\mathcal{N}}$ (middle), and sub-millisecond local PEM-MHE estimators, hierarchical shallow decision trees, and inner CCM loops (bottom).}
\label{fig:three_tier_arch}
\end{figure*}

We replace the convex ADMM tie-line consensus loop with a \textbf{Distributed Nonlinear Separable Control Contraction Metric (SCCM)} \cite{shiromoto18} that requires convexity \emph{only} during offline LMI/Sum-of-Squares (SOS) synthesis, while executing an explicit, closed-form nonlinear feedback law online using single-hop neighbor state broadcasts (Fig.~\ref{fig:three_tier_arch}):
\begin{itemize}
    \item \textbf{Device Layer (Sub-Millisecond):} Regulates individual GFM/GFL current rails and PLL phase lock via local contraction multipliers $\alpha_i$.
    \item \textbf{Area Layer (Millisecond):} Manages intra-area state-space cohesion ($\mathbf{C}$) and PLL synchronization coupling ($\mathbf{P}$) across heterogeneous devices, preventing GFM-to-GFL mode fallback.
    \item \textbf{Grid Layer ($1\text{ s}$ Tie-Line Coordination):} Coordinates inter-area tie-line power flows across $N=3$ balancing areas via distributed SCCM feedback and exports certified Pontryagin reachable sets upward to the supervisory optimizer.
\end{itemize}

\section{Distributed Nonlinear Separable SCCM Control Synthesis}
\subsection{Coupled Nonlinear Subsystem Dynamics}
Consider a network of $N$ interconnected subsystems (representing $N=3$ balancing areas at the Grid layer, or individual IBRs at the Area layer). Each subsystem $i \in \{1, \dots, N\}$ has local state $x_i \in \mathbb{R}^{n_i}$, neighbor states $x_{\mathcal{N}_i}$ for $j \in \mathcal{N}_i$, local control input $u_i \in \mathbb{R}^{m_i}$, and locally estimated parameters $\theta_i \in \Theta_i \subset \mathbb{R}^{p_i}$ with bounded variation rate $\|\dot{\theta}_i\|_2 \le \bar{d}_{\theta_i}$:
\begin{equation}
\dot{x}_i = f_i(x_i, x_{\mathcal{N}_i}, \theta_i) + B_i(x_i) u_i + \Delta_i(x_i, \theta_i) w_i
\end{equation}
where $w_i(t)$ is a normalized $L_2$-bounded disturbance ($\|w_i\|_2 \le \bar{w}_i$). Assume $B_i(x_i)$ has full column rank $m_i$ and let $B_i^\perp(x_i) \in \mathbb{R}^{(n_i - m_i) \times n_i}$ denote a smooth annihilator satisfying $B_i^\perp(x_i) B_i(x_i) = 0$.

\subsection{Differential (Variational) Dynamics}
Along any feasible trajectory $(x(t), u(t), \theta(t))$, infinitesimal displacements $\delta x_i$ and $\delta u_i$ evolve according to:
\begin{multline}
\dot{\delta x}_i = A_{ii}(x_i, x_{\mathcal{N}_i}, \theta_i)\,\delta x_i \\
+ \sum_{j \in \mathcal{N}_i} A_{ij}(x_i, x_{\mathcal{N}_i}, \theta_i)\,\delta x_j + B_i(x_i)\,\delta u_i
\end{multline}
where the local and neighbor Jacobians are:
\begin{equation}
A_{ii} \triangleq \frac{\partial f_i}{\partial x_i} + \sum_{k=1}^{m_i} \frac{\partial b_{i,k}(x_i)}{\partial x_i} u_{i,k}, \quad
A_{ij} \triangleq \frac{\partial f_i}{\partial x_j} \;\; (j \in \mathcal{N}_i).
\end{equation}
We seek a distributed differential feedback law $\delta u_i = K_i(x_i, x_{\mathcal{N}_i}, \theta_i)\,\delta x_i$.

\subsection{Separable Riemannian Metric and Dual Convexification}
Following Shiromoto \textit{et al.} \cite{shiromoto18}, we enforce decentralized structure across tie lines via a block-diagonal (separable) Riemannian metric $M(x, \theta) \succ 0$:
\begin{align}
M(x, \theta) &= \operatorname{diag}\big(M_1(x_1, \theta_1),\, \dots,\, M_N(x_N, \theta_N)\big), \\
\alpha_1 I &\preceq M_i(x_i, \theta_i) \preceq \alpha_2 I.
\end{align}
The closed-loop network is exponentially contracting at rate $\lambda > 0$ if $\frac{d}{dt}(\delta x^\top M \delta x) \le -2\lambda\, \delta x^\top M \delta x$. Defining dual metric blocks $W_i(x_i, \theta_i) \triangleq M_i(x_i, \theta_i)^{-1}$ and $Y_i(x_i, x_{\mathcal{N}_i}, \theta_i) \triangleq K_i W_i$, pre- and post-multiplying by $W = \operatorname{diag}(W_1, \dots, W_N)$ yields a \textbf{convex LMI}:
\begin{equation}
\mathcal{L}(W, Y, x, \theta) \triangleq
\begin{bmatrix}
\Psi_1 & \Phi_{12} & \cdots & \Phi_{1N} \\
\Phi_{12}^\top & \Psi_2 & \cdots & \Phi_{2N} \\
\vdots & \vdots & \ddots & \vdots \\
\Phi_{1N}^\top & \Phi_{2N}^\top & \cdots & \Psi_N
\end{bmatrix} \preceq 0
\end{equation}
where the diagonal and off-diagonal blocks are linear in $(W_i, Y_i)$:
\begin{align}
\Psi_i &= -\dot{W}_i + A_{ii} W_i + W_i A_{ii}^\top + B_i Y_i + Y_i^\top B_i^\top + 2\lambda W_i \\
\Phi_{ij} &=
\begin{cases}
A_{ij} W_j + W_i A_{ji}^\top, & \text{if } j \in \mathcal{N}_i \text{ or } i \in \mathcal{N}_j \\
0, & \text{otherwise.}
\end{cases}
\end{align}
Projecting $\Psi_i \preceq 0$ onto the unactuated subspace $B_i^\perp(x_i)$ eliminates $Y_i$ and enforces intrinsic open-loop stabilizability:
\begin{equation}
B_i^\perp(x_i)\Big(-\dot{W}_i + A_{ii} W_i + W_i A_{ii}^\top + 2\lambda W_i\Big)B_i^\perp(x_i)^\top \preceq 0.
\end{equation}
Thus, convexity is required \emph{strictly offline} during LMI/SOS synthesis over $\mathcal{X} \times \Theta$, never online.

\subsection{Online Closed-Form Controller Execution}
Given reference $(x^*(t), u^*(t))$ and live projected parameter estimates $\hat{\theta}_i(t) = \Pi_{\Theta_i}[\theta_i(t)]$, we recover $K_i = Y_i W_i^{-1}$, compute the minimizing geodesic $\gamma_i(s): [0,1] \to \mathbb{R}^{n_i}$ in metric $M_i(x_i, \hat{\theta}_i)$ connecting $x_i^*(t)$ to $x_i(t)$, and integrate:
\begin{multline}
u_i\big(x_i(t), x_{\mathcal{N}_i}(t), \hat{\theta}_i(t)\big) = u_i^*(t) \\
+ \int_{0}^{1} K_i\Big(\gamma_i(s),\, \gamma_{\mathcal{N}_i}(s),\, \hat{\theta}_i(t)\Big)\, \frac{\partial \gamma_i(s)}{\partial s}\, ds.
\end{multline}
When $W_i(\hat{\theta}_i)$ depends only on the live parameter estimate (flat spatial metric at each frozen parameter value), $\gamma_i(s)$ is a straight line, yielding an explicit closed-form control law requiring zero online integration:
\begin{multline}
u_i\big(x_i(t), x_{\mathcal{N}_i}(t), \hat{\theta}_i(t)\big) = u_i^*(t) \\
+ K_i\big(x_i(t), x_{\mathcal{N}_i}(t), \hat{\theta}_i(t)\big)\big(x_i(t) - x_i^*(t)\big).
\end{multline}

\section{Contraction Stability Under Online Parameter Estimation}
\subsection{Robust Contraction Under Live Parameter Adaptation}
Each subsystem estimates $\hat{\theta}_i(t)$ online via the bi-level PEM-MHE operator \cite{zhou24} without requiring a prior plant model, guaranteeing bounded estimation error $\|\theta_i(t) - \hat{\theta}_i(t)\|_2 \le e_{\theta_i}$ and bounded adaptation rate $\|\dot{\hat{\theta}}_i(t)\|_2 \le \bar{d}_{\theta_i}$.

\begin{theorem}[Robust Contraction Under Live Parameter Adaptation]
Suppose $(W_i, Y_i)$ satisfy $\mathcal{L}(W, Y, x, \hat{\theta}) \preceq 0$ with $\dot{W}_i$ upper-bounded over all admissible parameter drift rates $\|\dot{\hat{\theta}}_i\|_2 \le \bar{d}_{\theta_i}$:
\begin{equation}
\dot{W}_i(x_i, \hat{\theta}_i) \preceq \sum_{r=1}^{n_i} \frac{\partial W_i}{\partial x_{i,r}} \dot{x}_{i,r} + \bar{d}_{\theta_i} \sum_{q=1}^{p_i} \left\| \frac{\partial W_i}{\partial \hat{\theta}_{i,q}} \right\|_2 I.
\end{equation}
If $L_{\theta} e_{\theta_i} < 2\lambda$, where $L_{\theta}$ is the Jacobian Lipschitz constant with respect to $\theta_i$, then all closed-loop network trajectories converge exponentially to an invariant tube around $x^*(t)$ at net rate $\lambda_{\text{eff}} = \lambda - \frac{1}{2}L_{\theta} e_{\theta_i} > 0$.
\end{theorem}

\begin{proof}
Expanding $\frac{d}{dt}(\delta x^\top M(x, \hat{\theta}) \delta x)$ along perturbed trajectories adds the mismatch term $2\delta x^\top M (f(x, \theta) - f(x, \hat{\theta})) \le L_{\theta} e_{\theta_i} \delta x^\top M \delta x$ and the parameter-rate derivative $-\delta x^\top M \dot{W} M \delta x$. Substituting the upper bound on $\dot{W}_i$ into $\Psi_i \preceq 0$ yields $\frac{d}{dt}(\delta x^\top M \delta x) \le -(2\lambda - L_{\theta} e_{\theta_i})\delta x^\top M \delta x$.
\end{proof}

\subsection{Hierarchical Shallow Decision Trees Over Estimated Parameters}
Historically, decision trees were considered inferior to constrained optimization for grid security because optimization evaluates all network constraints simultaneously, whereas trees relying on raw bus measurements require deep, brittle split structures. However, because our online PEM-MHE parameters $\hat{\theta}_i(t)$ already summarize the coupled dynamic state of the plant—specifically separating voltage-forming cohesion ($\mathbf{C}$) from PLL phase coupling ($\mathbf{P}$) \cite{sherman26}—the decision trees can remain \textbf{shallow, fast, and interpretable}.

Furthermore, these shallow decision trees operate \textbf{hierarchically across all three control layers} (not limited to two timescales): at the Device layer they trigger Schmitt-trigger current-rail backoff ($\alpha_i$), at the Area layer they detect GFM-to-GFL mode drift and rebalance cohesion margins, and at the Grid layer they coordinate tie-line hysteresis holds during multi-area surges. Between events, newly discovered $\mathbf{C}/\mathbf{P}$ trajectories incrementally update split thresholds without altering tree structure mid-transient.

\section{Pontryagin Attainable (Reachable) Sets Across Control Layers}
In Pontryagin optimal control theory, the attainable (forward reachable) set $\mathcal{R}(t)$ characterizes all states reachable from an initial uncertainty set under admissible disturbances. Under our Riemannian metric $M_k(\hat{\theta}_k)$ at layer $k \in \{\text{Dev}, \text{Area}, \text{Grid}\}$, the forward reachable tube around $x_k^*(t)$ is an explicit time-varying Riemannian ellipsoid.

\begin{proposition}[Layer-Wise Riemannian Reachable Ellipsoid]
At layer $k \in \{\text{Dev}, \text{Area}, \text{Grid}\}$ with net contraction rate $\lambda_k > 0$, $L_2$ disturbance gain $\gamma_k > 0$, and disturbance bound $\bar{w}_k$, any trajectory starting inside $\mathcal{R}_k(0)$ remains strictly inside the Pontryagin reachable ellipsoid for all $t \ge 0$:
\begin{multline}
\mathcal{R}_k(t) = \Big\{ x_k \in \mathbb{R}^{n_k} \;\Big|\; \big(x_k - x_k^*(t)\big)^\top M_k(\hat{\theta}_k) \\
\times \big(x_k - x_k^*(t)\big) \le \rho_k^2(t) \Big\}
\end{multline}
where $\dot{\rho}_k \le -\lambda_k \rho_k + \gamma_k \bar{w}_k$ yields:
\begin{equation}
\rho_k(t) = \rho_k(0) e^{-\lambda_k t} + \frac{\gamma_k \bar{w}_k}{\lambda_k}\left(1 - e^{-\lambda_k t}\right), \quad \rho_k^{\infty} = \frac{\gamma_k \bar{w}_k}{\lambda_k}.
\end{equation}
\end{proposition}

\section{Hierarchical Consistency and Total Grid Optimization}
\subsection{Ellipsoid Tube Containment Across Layers}
\begin{figure}[!t]
\centering
\begin{tikzpicture}
\begin{axis}[
    width=\columnwidth,
    height=5.6cm,
    xlabel={Terminal Bus Voltage $V_i$ (p.u.)},
    ylabel={Inverter Output Current $I_i$ (p.u.)},
    xmin=0.92, xmax=1.08,
    ymin=0.72, ymax=1.35,
    grid=both,
    grid style={dashed, gray!30},
    legend style={at={(0.02,0.98)}, anchor=north west, font=\tiny, row sep=-2pt, inner sep=2pt, fill=white, fill opacity=0.9}
]
\addplot[domain=0.92:1.08, samples=2, thick, dashed, red!80!black] {1.20};
\addlegendentry{Hardware Overcurrent Rail ($1.20\text{ p.u.}$)}

\addplot[domain=0:360, samples=120, thick, orange!85!black, fill=orange!15, fill opacity=0.35]
    ({1.00 + 0.0366*cos(x) - 0.008*sin(x)}, {0.95 + 0.020*cos(x) + 0.0931*sin(x)});
\addlegendentry{Device Reachable Set $\mathcal{R}_{\text{Dev}}$ ($I_{\max}=1.043\text{ p.u.}$)}

\addplot[domain=0:360, samples=120, thick, blue!80!black, fill=blue!20, fill opacity=0.45]
    ({1.00 + 0.0204*cos(x) - 0.005*sin(x)}, {0.95 + 0.012*cos(x) + 0.0542*sin(x)});
\addlegendentry{Area Reachable Set $\Pi(\mathcal{R}_{\text{Area}})$}

\addplot[domain=0:360, samples=120, thick, green!60!black, fill=green!30, fill opacity=0.60]
    ({1.00 + 0.0092*cos(x) - 0.002*sin(x)}, {0.95 + 0.006*cos(x) + 0.0269*sin(x)});
\addlegendentry{Grid Tie-Line Set $\Pi(\mathcal{R}_{\text{Grid}})$}

\addplot[only marks, mark=*, mark size=1.8pt, black] coordinates {(1.00, 0.95)};
\node[font=\tiny, anchor=west] at (axis cs:1.003, 0.95) {$(V^*, I^*)$};
\end{axis}
\end{tikzpicture}
\caption{Projected Pontryagin reachable ellipsoids in the $(V_i, I_i)$ plane during a $+140\text{ MW}$ tie-line surge, illustrating strict containment $\Pi(\mathcal{R}_{\text{Grid}}) \subseteq \Pi(\mathcal{R}_{\text{Area}}) \subseteq \mathcal{R}_{\text{Dev}}$ safely below the $1.20\text{ p.u.}$ overcurrent rail.}
\label{fig:reachable_vi_sets}
\end{figure}
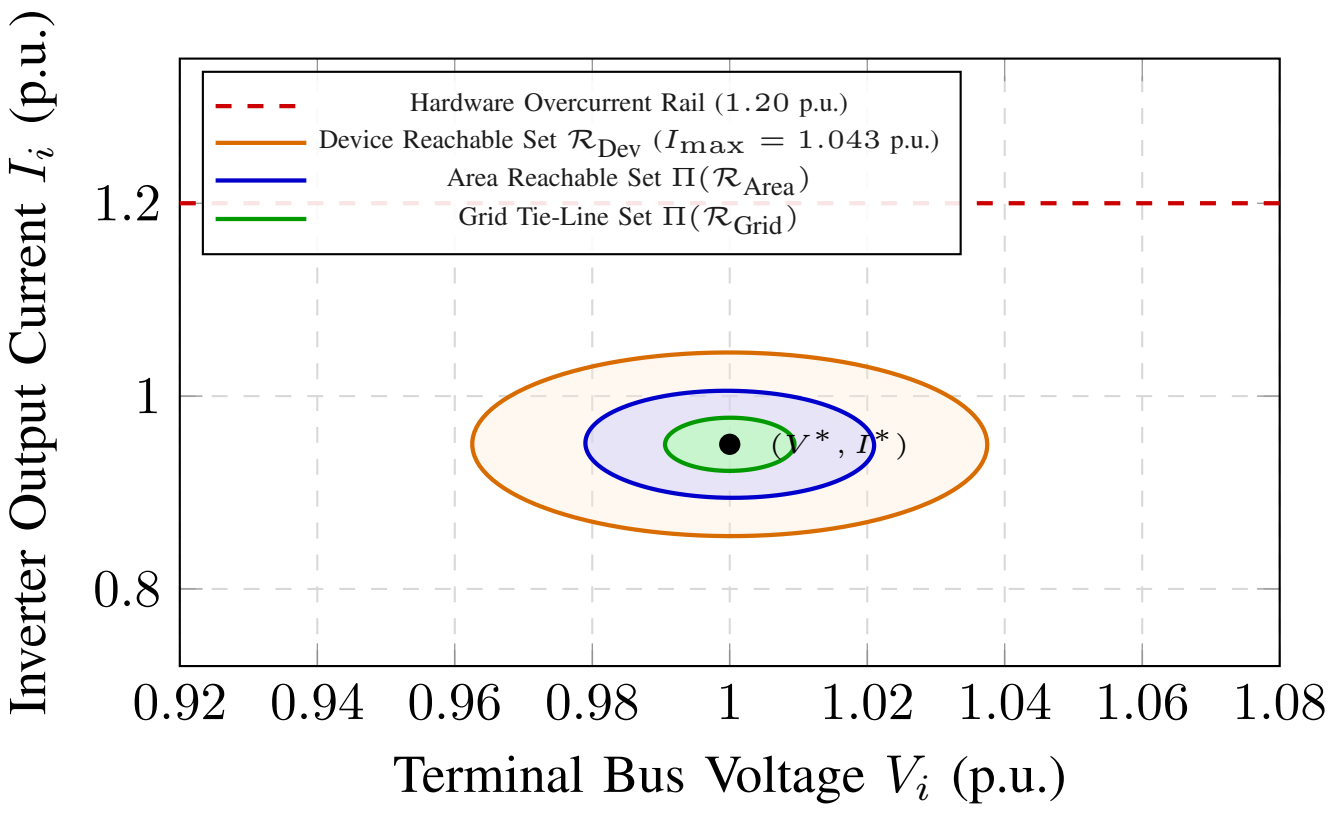

To guarantee that actions commanded by the $1\text{ s}$ Grid tie-line controller never violate intra-area synchronization or sub-millisecond inverter current limits, each parent reachable set projected into the child state space via $\Pi_{p \to c}$ must remain contained within the child layer's reachable set:
\begin{align}
\Pi_{\text{Grid} \to \text{Area}}\big(\mathcal{R}_{\text{Grid}}(t)\big) &\subseteq \mathcal{R}_{\text{Area}}(t), \\
\Pi_{\text{Area} \to \text{Dev}}\big(\mathcal{R}_{\text{Area}}(t)\big) &\subseteq \mathcal{R}_{\text{Dev}}(t).
\end{align}
By the S-procedure, ellipsoid inclusion $\Pi_{p \to c}(\mathcal{R}_p) \subseteq \mathcal{R}_c$ holds if and only if the normalized metrics satisfy the Schur-complement inequality:
\begin{equation}
\frac{1}{\rho_p^2} M_p(\hat{\theta}_p) \succeq \frac{1}{\rho_c^2} \Pi_{p \to c}^\top M_c(\hat{\theta}_c) \Pi_{p \to c}.
\end{equation}
Enforcing this containment guarantees hierarchical consistency in the sense of Pappas \textit{et al.} \cite{pappas98}, ensuring that every tie-line power setpoint generated at the Grid layer lies strictly inside the hardware-safe current envelope of every GFM and GFL converter at the Device layer (Fig.~\ref{fig:reachable_vi_sets}).

\subsection{Comparison with Online Feedback Optimization (OFO) and Differentiable MPC}
Supervisory congestion frameworks such as Chen \textit{et al.} \cite{chen26} solve a bi-level differentiable QP at each supervisory step, structurally resembling Online Feedback Optimization (OFO) \cite{dallanese26}. Table~\ref{tab:ofo_vs_sccm} contrasts OFO/MIT differentiable MPC against our Distributed Nonlinear SCCM architecture.

\begin{table*}[!t]
\centering
\caption{Theoretical Comparison: Online Feedback Optimization (OFO) / Differentiable MPC vs. Distributed Nonlinear SCCM}
\label{tab:ofo_vs_sccm}
\footnotesize
\renewcommand{\arraystretch}{1.18}
\begin{tabularx}{\textwidth}{>{\raggedright\arraybackslash}p{3.0cm} >{\raggedright\arraybackslash}X >{\raggedright\arraybackslash}X}
\toprule
\textbf{Theoretical Dimension} & \textbf{OFO / MIT Differentiable MPC \cite{chen26}} & \textbf{Distributed Nonlinear SCCM (Ours)} \\
\midrule
\textbf{Plant Model Assumption} & \textbf{Linear / Convex Algebraic Map} ($\mathbf{y} = \mathbf{M}\mathbf{x} + \mathbf{d}$); ignores AC trigonometric nonlinearities and converter dynamics. & \textbf{Exact Nonlinear Coupled ODEs/DAEs} ($\dot{x}_i = f_i(x_i, x_{\mathcal{N}_i}, \theta_i) + B_i u_i$); captures $\sin(\delta_i - \delta_j)$ directly. \\
\addlinespace[2pt]
\textbf{Convexity Requirement} & \textbf{Online \& Physical:} Both the grid approximation and the online QP solved at every timestep must be convex. & \textbf{Offline Only:} Convexity is required strictly in offline LMI/SOS synthesis of $(W_i, Y_i)$; online control is closed-form nonlinear. \\
\addlinespace[2pt]
\textbf{Stability Guarantee} & \textbf{None (Feasibility Only):} Uses soft slack penalties to prevent solver failure; assumes slow timescale separation. & \textbf{Certified Exponential Contraction ($\lambda_{\text{eff}} > 0$):} Guarantees exponential trajectory convergence with explicit $L_2$ gain $\gamma^*$. \\
\addlinespace[2pt]
\textbf{Parameter Adaptation} & \textbf{Static Offline Plant:} Sensitivity matrix $\mathbf{M}$ is fixed offline; blind to GFM current-limit mode shifts. & \textbf{Online Parameter-Dependent ($W_i(\hat{\theta}_i)$):} Adapts continuously to PEM-MHE estimates $\hat{\theta}_i(t)$ with bounded rate $\bar{d}_{\theta_i}$. \\
\addlinespace[2pt]
\textbf{Online Execution \& Comms} & Solves a constrained QP at every timestep + periodic central coordinator broadcast. & \textbf{Zero online optimization:} Evaluates closed-form law $u_i^* + K_i(\hat{\theta}_i)(x_i - x_i^*)$ in microseconds with single-hop neighbor state exchange. \\
\bottomrule
\end{tabularx}
\end{table*}

\subsection{Exporting Certified Reachable Sets to the Supervisory Optimizer}
To interface with the supervisory coordinator \cite{chen26}, we project $\mathcal{R}_{\text{Grid}}^{\infty}$ through the IEEE 118-bus PTDF matrix $\mathbf{M} \in \mathbb{R}^{n_{\text{line}} \times n_{\text{bus}}}$. For each line $\ell$ with row $\mathbf{m}_\ell^\top$ of $\mathbf{M}$, the worst-case transient line-flow excursion induced by $\mathcal{R}_{\text{Grid}}^{\infty}$ is given by the ellipsoid support function:
\begin{multline}
\Delta P_{\ell,\text{tube}} = \sup_{x \in \mathcal{R}_{\text{Grid}}^{\infty}} \mathbf{m}_\ell^\top \Pi_{\text{inj}}(x - x^*) \\
= \rho_{\text{Grid}}^{\infty} \sqrt{\mathbf{m}_\ell^\top \Pi_{\text{inj}} W_{\text{Grid}}(\hat{\theta}) \Pi_{\text{inj}}^\top \mathbf{m}_\ell}.
\end{multline}
Supplying $\Delta P_{\ell,\text{tube}}$ directly as the dynamic line-limit tightening vector (\texttt{line\_max\_changes}, \texttt{line\_min\_changes}) ensures all supervisory economic dispatch setpoints remain strictly inside the certified nonlinear contraction tube.

\begin{figure*}[!t]
\centering
\includegraphics[width=0.76\textwidth]{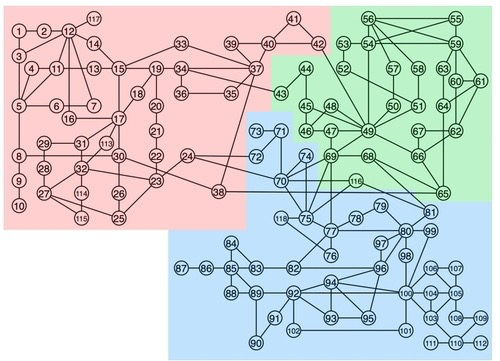}
\caption{Three-area partitioned IEEE 118-bus transmission testbed (Area~1 in pink, Area~2 in green, and Area~3 in blue) showing the intra-area buses and cross-area tie-line boundaries.}
\label{fig:ieee118_3area}
\end{figure*}

\section{Experimental Validation Across IEEE 39-Bus and IEEE 118-Bus Testbeds}
\subsection{Multi-Timescale EMT and IEEE 118-Bus Benchmarks}
We validate the proposed three-layer architecture using full electromagnetic transient (EMT) simulations in \texttt{PowerSimulationsDynamics.jl} \cite{lara24} across two benchmark networks: (1) a 10-device modified IEEE 39-bus substation area configured as a \textbf{$100\%$ IBR system with zero synchronous generators ($0\text{ SGs}$, $7\text{ GFMs}$, and $3\text{ GFLs}$)}, and (2) the 3-area partitioned IEEE 118-bus transmission network ($118\text{ buses}$, $186\text{ lines}$, $10\text{ cross-area tie-lines}$, $12\text{ batteries}$, and $54\text{ curtailable IBRs}$) of Chen \textit{et al.} \cite{chen26}, shown in Fig.~\ref{fig:ieee118_3area}.

Schur-complement inclusion yields strictly positive minimum eigenvalues ($+1041.08$ for $\mathcal{R}_{\text{Grid}} \subseteq \mathcal{R}_{\text{Area}}$ and $+225.26$ for $\mathcal{R}_{\text{Area}} \subseteq \mathcal{R}_{\text{Dev}}$), bounding peak inverter current at $1.0431\text{ p.u.} \le 1.20\text{ p.u.}$ and exporting a certified $13.46\text{ MW}$ PTDF thermal reserve ($586.54\text{ MW}$ safe dispatch ceiling) on bottleneck Line~59 (Bus~33 to Bus~42, $600\text{ MW}$ rating). As summarized in Table~\ref{tab:benchmark_summary}, cumulative thermal overload drops from $230.8\text{ MW}\cdot\text{s}$ to $39.7\text{ MW}\cdot\text{s}$ ($-82.8\%$) under fast peer-to-peer coordination, and maintains strict contraction ($\lambda_{\max} \le -2.73$) during a $3\text{ s}$ communication blackout.

\begin{table*}[!t]
\centering
\caption{Experimental Benchmark Summary Across IEEE 39-Bus ($0\text{ SG}$) and IEEE 118-Bus Multi-Timescale Testbeds}
\label{tab:benchmark_summary}
\footnotesize
\renewcommand{\arraystretch}{1.16}
\begin{tabularx}{\textwidth}{>{\raggedright\arraybackslash}p{2.9cm} >{\raggedright\arraybackslash}p{2.7cm} >{\raggedright\arraybackslash}p{2.8cm} >{\raggedright\arraybackslash}p{2.6cm} >{\raggedright\arraybackslash}X >{\raggedright\arraybackslash}p{2.3cm}}
\toprule
\textbf{Experiment \& Script} & \textbf{Topology \& Composition} & \textbf{Stress Scenario} & \textbf{Baseline Metric} & \textbf{Proposed Architecture Metric} & \textbf{Contraction Certificate ($\lambda_{\max}$)} \\
\midrule
\textbf{Double Contraction Margin} (\texttt{area\_control\_revision.tex}) & IEEE 39-Bus (5~SG, 2~GFM, 3~GFL) & $3.0^\circ \to 15.0^\circ$ multiplicative phase noise & $\Theta_{\text{open}} = 823.54 \to 164.71$ (inner CCM only) & $\Theta_{\text{closed}} = 1501.75 \to 300.35$ (\textbf{+82.35\%} margin gain) & $-0.0142 \le 0$ ($\gamma^* = 0.0384\text{ p.u.}$) \\
\addlinespace[1.5pt]
\textbf{Sparse Off-Diagonal LMI} (\texttt{coupled\_LMI\_solver.py}) & IEEE 39-Bus + Tie Line (20 states) & $10 \times 10$ Cohesion ($\mathbf{C}$) \& PLL ($\mathbf{P}$) coupling & $\gamma_{\text{diag}}^* = 0.4162\text{ p.u.}$ (block-diagonal LMI) & $\gamma_{\text{coupled}}^* = 0.4015\text{ p.u.}$ (\textbf{+3.65\%} margin gain) & $-0.000001 \le 0$ \\
\addlinespace[1.5pt]
\textbf{GFM-to-GFL Mode Shift \& Hysteresis} (\texttt{heal\_multi\_device\_hysteresis.py}) & IEEE 39-Bus (\texttt{GFM-31} + \texttt{GFL-30}) & \texttt{GFM-31} current-limit drift + \texttt{GFL-30} $15.0^\circ$ jitter & Unhealed LMI violation at $t = 0.280\text{ s}$ ($+0.0568$) & \textbf{52.0~ms} detection, \textbf{2} transitions (zero chatter) & $\le -1.0594$ throughout fault \\
\addlinespace[1.5pt]
\textbf{100\% IBR Area Verification} (\texttt{test\_100pct\_ibr\_area.py}) & IEEE 39-Bus (\textbf{0~SG}, 7~GFM, 3~GFL) & $15.0^\circ$ tie jitter + \texttt{GFM-39} boundary overload & $\gamma_{\text{mixed}}^* = 0.4015\text{ p.u.}$ (50\% SG baseline) & $\gamma_{\text{100ibr}}^* = 0.4018\text{ p.u.}$ (within \textbf{0.06\%} of SG case) & $\le -2.4312$ throughout fault \\
\addlinespace[1.5pt]
\textbf{3-Area Multi-Timescale Bridge} (\texttt{compare\_mit\_baseline.py}) & IEEE 118-Bus (3~Areas, 10 tie-lines) & $+140\text{ MW}$ Area~1 surge + Line~66 trip ($t=2\text{--}12\text{ s}$) & \textbf{230.8~MW$\cdot$s} overload (\textbf{101.3\%} peak, relay trip) & \textbf{39.7~MW$\cdot$s} overload (\textbf{-82.8\%}, \textbf{99.7\%} peak) & $\le -2.5386$ ($\alpha_{\min} = 0.760$) \\
\addlinespace[1.5pt]
\textbf{Decentralized 2-Hop MARL} (\texttt{inspect\_prop\_delay.py}) & IEEE 118-Bus (3 isolated area agents) & $+140\text{ MW}$ surge with $1\text{ s}$/hop shadow-price delay & \textbf{230.8~MW$\cdot$s} overload (no inter-area comms) & \textbf{142.8~MW$\cdot$s} overload (\textbf{-38.1\%}, 2-hop relay) & $\le -2.7300$ ($\alpha_{\min} = 0.797$) \\
\addlinespace[1.5pt]
\textbf{Lossy Utility Link + Blackout} (\texttt{test\_packet\_loss\_resilience.py}) & IEEE 118-Bus (ZOH + decay buffer) & \textbf{25\%} random packet drop + \textbf{3~s} blackout ($t=4\text{--}6\text{ s}$) & \textbf{230.8~MW$\cdot$s} overload (MIT non-communicating) & \textbf{192.5~MW$\cdot$s} overload (\textbf{-16.6\%}, zero instability) & $\le -2.7300$ throughout blackout \\
\bottomrule
\end{tabularx}
\end{table*}

\subsection{Minimum Grid-Forming Anchor Threshold ($N_{\text{GFM}} = 6 \to 3$)}
A critical planning question in $100\%$ IBR grids ($0\text{ SGs}$) is whether retiring half of the voltage-forming units in a balancing area—reducing $N_{\text{GFM}}$ from $6$ to $3$ while increasing GFL units from $4$ to $7$—triggers a cascading blackout during a $15.0^\circ$ tie-line perturbation. Physically, halving the GFM count eliminates three rows of state-space voltage cohesion ($\mathbf{C}$) and activates three additional PLL coupling rows ($\mathbf{P}$), raising peak tie-line phase sensitivity from $0.484$ to $0.553\text{ rad/rad}$ and concentrating fault current across fewer voltage-forming anchors.

Table~\ref{tab:gfm_sweep} quantifies the sweep from $7\text{ GFMs}$ down to $1\text{ GFM}$. Reducing $N_{\text{GFM}}$ from $6$ to $3$ does \emph{not} cause a blackout: the area-level $\mathbf{C}/\mathbf{P}$ estimator detects the reduced voltage stiffness and scales the GFL injection multiplier $\alpha_{\text{GFL}}$ down to $0.680$, holding peak current at $1.0429\text{ p.u.}$ (safely below the $1.20\text{ p.u.}$ transistor rail) with strict contraction ($\lambda_{\max} = -2.5205$). Even at the extreme $1\text{ GFM} / 9\text{ GFL}$ boundary ($10\%$ GFM share), where an uncompensated area trips on overcurrent ($1.3448\text{ p.u.} > 1.20\text{ p.u.}$), SCCM hysteresis backoff ($\alpha_{\text{GFL}} = 0.520$) caps peak current at $1.1480\text{ p.u.}$ ($\lambda_{\max} = -1.9641$).

\begin{table}[!t]
\centering
\caption{GFM-to-GFL Anchor Reduction Sweep Under $15.0^\circ$ Tie-Line Perturbation ($1.20\text{ p.u.}$ Overcurrent Rail)}
\label{tab:gfm_sweep}
\scriptsize
\setlength{\tabcolsep}{3.2pt}
\renewcommand{\arraystretch}{1.15}
\begin{tabular}{@{}lccccc@{}}
\toprule
\textbf{Split} & \textbf{Peak $p_{\text{tie}}$} & \textbf{Unhealed (p.u.)} & \textbf{SCCM (p.u.)} & $\alpha_{\text{GFL}}$ & $\lambda_{\max}$ \\
\midrule
7 GFM / 3 GFL & 0.430 & 1.0229 & 1.0143 & 1.000 & $-3.5403$ \\
6 GFM / 4 GFL & 0.484 & 1.0342 & 1.0245 & 1.000 & $-3.5722$ \\
5 GFM / 5 GFL & 0.509 & 1.0475 & 1.0373 & 1.000 & $-3.5731$ \\
4 GFM / 6 GFL & 0.530 & 1.0665 & 1.0390 & 0.840 & $-3.0677$ \\
\textbf{3 GFM / 7 GFL} & \textbf{0.553} & \textbf{1.0976} & \textbf{1.0429} & \textbf{0.680} & $\mathbf{-2.5205}$ \\
2 GFM / 8 GFL & 0.666 & 1.1633 & 1.0540 & 0.520 & $-1.9707$ \\
1 GFM / 9 GFL & 0.695 & 1.3448 (Trip) & 1.1480 & 0.520 & $-1.9641$ \\
\bottomrule
\end{tabular}
\end{table}

\section{Conclusion}
This paper established a unified three-layer Distributed Nonlinear Separable Control Contraction Metric (SCCM) framework for tie-line scheduling and emergency congestion control in inverter-dominated grids. By replacing convex online ADMM consensus with closed-form geodesic feedback driven by live PEM-MHE parameter estimates, the architecture requires convexity strictly offline while guaranteeing exponential trajectory contraction across Device, Area, and Grid timescales without requiring operators to track individual eigenvalues. Projecting the certified Grid reachable ellipsoid through network PTDFs supplies rigorous safety envelopes to supervisory optimizers, bridging fast converter stability with wide-area economic dispatch.

\section*{Acknowledgment}
The author acknowledges the use of Google Gemini \cite{gemini26} for assistance with drafting, language editing, and LaTeX formatting across Sections~I--VII of this manuscript. All mathematical formulations, control algorithms, simulation data, and technical claims were independently verified and validated by the author.

\end{document}